\documentclass{vgtc}                          

\ifpdf
  \pdfoutput=1\relax                   
  \usepackage{graphicx}                
  \DeclareGraphicsExtensions{.pdf,.png,.jpg,.jpeg} 
\else
  \usepackage{graphicx}                
  \DeclareGraphicsExtensions{.eps}     
\fi%

\graphicspath{{figures/}{pictures/}{images/}{./}} 

\usepackage{microtype}                 
\PassOptionsToPackage{warn}{textcomp}  
\usepackage{textcomp}                  
\usepackage{mathptmx}                  
\usepackage{times}                     
\usepackage{cite}                      
\usepackage{tabu}                      
\usepackage{booktabs}                  
\usepackage{balance}

\onlineid{1231}

\vgtccategory{Research}

\vgtcinsertpkg

\title{Exploration of Hands-free Text Entry Techniques For Virtual Reality}

\author{Xueshi Lu\\ %
        \scriptsize Xi'an Jiaotong-Liverpool University %
\and Difeng Yu\\ %
     \scriptsize The University of Melbourne %
\and Hai-Ning Liang\thanks{Corresponding author: haining.liang@xjtlu.edu.cn}\\ %
    \scriptsize Xi'an Jiaotong-Liverpool University %
\and Wenge Xu\\ %
    \scriptsize Xi'an Jiaotong-Liverpool University %
\and Yuzheng Chen\\ %
    \scriptsize Xi'an Jiaotong-Liverpool University %
\and Xiang Li\\ %
    \scriptsize Xi'an Jiaotong-Liverpool University %
\and Khalad Hasan\\ %
    \scriptsize  University of British Columbia - Okanagan%
     }

\abstract{Text entry is a common activity in virtual reality (VR) systems. There is a limited number of available hands-free techniques, which allow users to carry out text entry when users’ hands are busy such as holding items or hand-based devices are not available. The most used hands-free text entry technique is DwellType, where a user selects a letter by dwelling over it for a specific period. However, its performance is limited due to the fixed dwell time for each character selection. In this paper, we explore two other hands-free text entry mechanisms in VR: BlinkType and NeckType, which leverage users’ eye blinks and neck’s forward and backward movements to select letters. With a user study, we compare the performance of the two techniques with DwellType. Results show that users can achieve an average text entry rate of 13.47, 11.18 and 11.65 words per minute with BlinkType, NeckType, and DwellType, respectively. Users’ subjective feedback shows BlinkType as the preferred technique for text entry in VR.%
} 


\CCScatlist{
  \CCScatTwelve{Human-centered computing}{Human computer interaction (HCI)}{Interaction paradigms}{Virtual reality};
  \CCScatTwelve{Human-centered computing}{Human computer interaction (HCI)}{Interaction techniques}{Text input}
}

\begin{document}


\firstsection{Introduction}

\maketitle
Text entry is essential for all types of interactive systems, including Virtual Reality (VR). Users, for example, often need to input passwords to login to a system or send text messages to chat with other users in a virtual social environment. Typically, text entry in VR is commonly performed through hand-held controllers \cite{speicher2018selection,yu2018pizzatext}. However, there are situations where users’ hands are occupied and cannot use their hands to hold a controller to perform text entry \cite{liang2019evaluating}; alternatively, oftentimes a controller or other hand-based devices are not readily available. For example, a doctor who is using a headset for surgery training might not be able to use a controller to perform text entry as his/her hands are occupied with holding surgical tools. In addition, users who have hand/arm disabilities are likely not able to use their hands to hold a controller \cite{xu2019dmove}. In these scenarios, an efficient and usable hands-free text entry technique would be the most convenient solution.\par

\begin{figure}[tb]
 \centering 
 \includegraphics[width=\columnwidth]{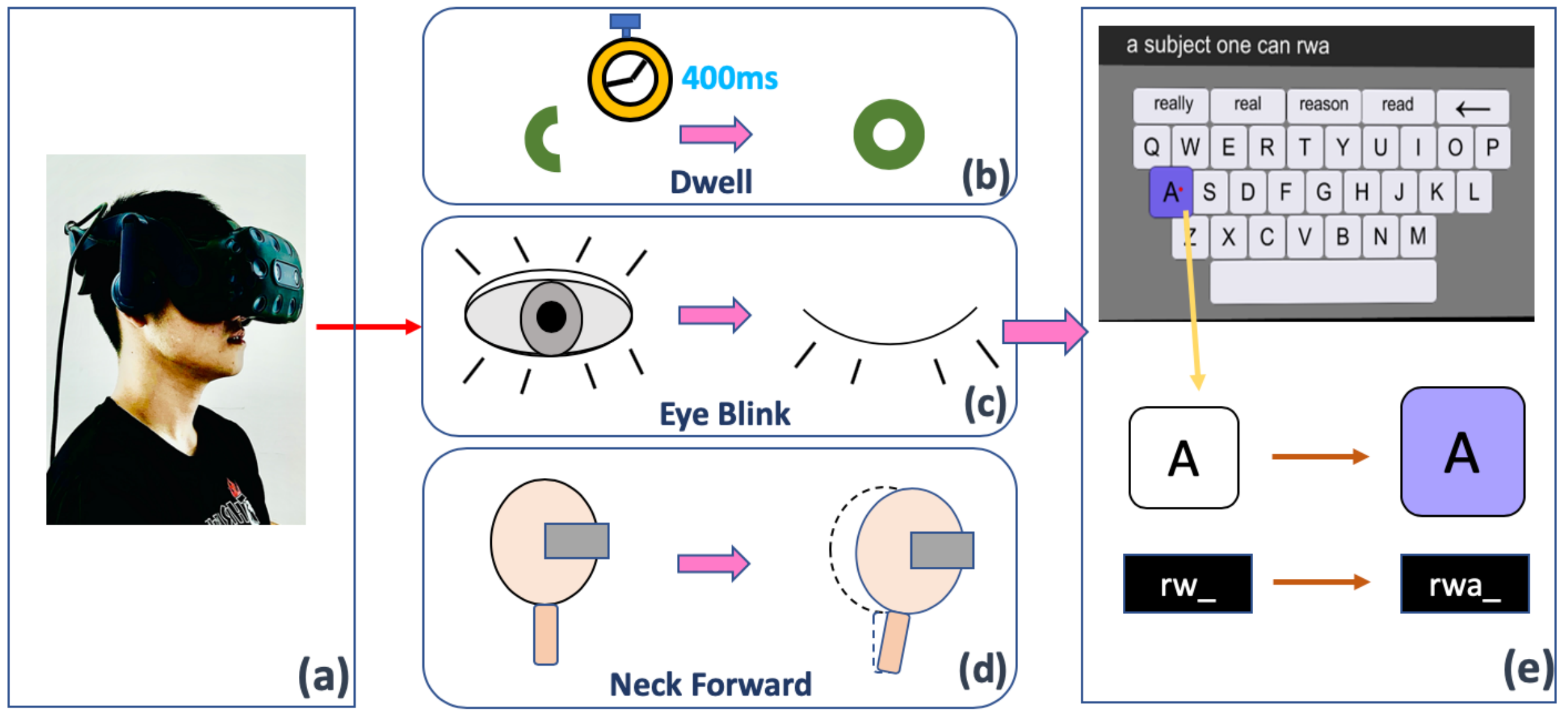}
 \caption{An illustration of the text entry with three hands-free techniques in VR. (a) A user wearing a head-mounted device can use (b) DwellType where a selection is done with by dwelling over a key; (c) BlinkType that where eye blinking is used to trigger a selection; and (d) NeckType, where neck movement is used, invoke a key. (e) After a successful selection, the target key becomes larger and highlighted.}
 \label{fig:sample}
\end{figure}

A limited number of hands-free text entry techniques have been proposed for VR. One of the most commonly used hands-free technique in VR is the dwell-based technique, where users enter text by controlling a virtual pointer with head motions and keep the pointer hovering on a target letter for a certain period of time to trigger a selection \cite{yu2017tap, xu2019pointing, yu2019modeling}. However, the dwell-based technique has certain limitations such as: (1) a long dwell time may decrease performance but a short dwell time can cause false-positive selections and errors \cite{jacob1991use}; (2) a pre-set dwell time always “pushed” users to select a target key and quickly move to the next one, a process that can be stressful \cite{kristensson2012potential}; and (3) keeping the pointer static for a while to avoid selecting unintentional keys could further lead to eye and neck fatigue \cite{sarcar2013eyek}. In addition to dwelling, speech-based techniques have been proposed as viable solutions to allow hands-free text entry \cite{pick2016swifter}. However, its performance suffers in noisy environments and could trigger privacy-related concerns \cite{xu2019ringtext}. Gaze-based text entry is commonly used in other settings, such as a common display (i.e., EyeK \cite{sarcar2013eyek}). However, eye-tracking systems are sensitive and not very reliable to enable gaze to be used for text entry that is efficient and have low error rates \cite{majaranta2002twenty}. In addition, gaze-based techniques often require rapid eye movements coupled with a period for the eye to be fixed and looking at one place (i.e., similar to a dwell time). Both activities can cause visual fatigue due to the repetitive nature of text entry \cite{dube2019text}.\par
In this paper, we present two alternative hands-free text entry techniques for VR systems: BlinkType and NeckType (see Figure 1). BlinkType is an eye blink-based technique that allows users to select letters in VR by using their eye blinks (Figure 1c). NeckType allows users to use their neck’s forward and backward motion as the selection mechanism (Figure 1d). Prior work showed that eye blinking can also be used for hands-free text entry, where the blinking is used for letter selection \cite{xu2019ringtext}. However, eye blinking was evaluated on different platforms other than VR and for individuals with severe motor impairments. In addition, research has shown that it is still unclear how well-received eye blinking, as a mechanism for selection confirmation, is by users \cite{xu2019ringtext,mackenzie2011blinkwrite}.\par
We conducted a user study with 36 participants to evaluate the performance of these two techniques and compared it against DwellType. Our results showed that BlinkType, NeckType, and DwellType could achieve an average text entry rate of 13.47, 11.18 and 11.65 words per minute (wpm). Results also revealed that BlinkType is the most preferred technique for hands-free text entry in VR.\par
The main contribution of the paper is a comparative evaluation of three hands-free text entry techniques based on text entry performance, user experience, and stimulator sickness.

\section{Related Work}

While several approaches have been proposed for entering text in VR systems (see recent reviews in \cite{pick2016swifter, ruan2018comparing, majaranta2009fast}), few of them have explored and compared hands-free text entry techniques in depth. In this section, we review previous research related to text entry in VR and non-VR systems: hands-free typing and dwell-free typing. Moreover, we give a description of eye blink and its properties.

\subsection{Hands-free Typing}

Speech-based is one feasible method to achieve hands-free text entry. Pick et al. \cite{pick2016swifter} devised SWIFTER, a speech-based multimodal text entry system, which could lead to a text entry rate of 23.6 wpm in average. Furthermore, Ruan et al. \cite{ruan2018comparing} evaluated the state-of-art speech recognition using Baidu’s Deep Speech 2 and compared it to mobile touch-based keyboards for both English and Mandarin Chinese. Their results revealed that speech is nearly three times faster than keyboard. However, it also led to more errors. There are two disadvantages of speech techniques: (1) they need a relatively quiet environment; and (2) they can lead to privacy problems. These issues prevent speech text entry to be used in many public spaces like libraries, restaurants, shops, malls, etc.\par
Gaze typing, as suggested by studies in non-VR settings \cite{majaranta2009fast}, could be another possibility to allow hands-free text entry in VR. The most common way of gaze typing is using a dwell time, which requires a user to hover the cursor on the target key for a predetermined duration to trigger selection. The dwell time is usually set as a constant value between 400ms-1000ms and leads to a text entry speed from 5-10 wpm \cite{porta2008eye, gugenheimer2017facedisplay, grubert2018text}. To enable fast gaze typing, Majaranta et al. \cite{majaranta2009fast} allowed the users to personalize the dwell time. Personalization is often beneficial since users might have their preferred dwell time, especially during the different learning stages and environment settings. Rajanna et al. explored dwell typing while sitting and biking. They observed that sitting+dwell and biking+dwell achieved 9.36 and 8.07 wpm, respectively \cite{rajanna2018gaze}. In this research, we evaluated and compared gaze- and head-based methods in VR settings.

\subsection{Dwell-free Typing}

Apart from dwell-based techniques, researchers have proposed dwell-free approaches. A certain number of dwell-free text entry techniques rely on tracking users’ eyes. Kristensson et al. \cite{kristensson2012potential} explored the potential speed of dwell-free eye-typing. Their results show that users could reach a theoretical average input rate of 46 wpm after 40 minutes—that is, this approach could be twice as fast as gaze typing techniques.\par
Gaze-typing techniques using gesture-based approaches have also been studied \cite{xu2019ringtext}. EyeWriter \cite{wobbrock2007not}, a technique based on the gestural unistroke alphabet from EdgeWrite, is a gesture-based eye typing technique. Eye-S \cite{porta2008eye} is another type of gesture-based typing method that requires users to draw letters on 9 specific hotspots and can lead to a mean rate of 6.8 wpm. EyeSwipe \cite{kurauchi2016eyeswipe} allows users gaze-type the first and last characters of the words by reverse crossing and glancing through the vicinity of the middle characters, then displaying the candidate words. Users could reach 11.7 wpm after 30 minutes of practice using this system. \par
Besides gesture-based techniques, selection-based typing techniques are also widely applied to dwell-free typing techniques. EyeK \cite{sarcar2013eyek} allows users to type by moving the cursor inside-outside-inside the specific area. In Filteryedping \cite{pedrosa2015filteryedping}, the user looks at characters in the same order of the required word, then the system automatically filters out unwanted words while and ranks the candidate words based on the length and frequency. This technique is claimed to reach a mean text entry rate of 15.95 wpm. Blink is a viable way to control input as well. BlinkWrite2 \cite{ashtiani2010blinkwrite2} is an eye blink text entry system, where participants could reach an average text entry speed of 5.3 wpm. BlinkWrite2 presents the potential of blink typing. However, the technique has not been tested with VR HMDs and its performance is low compared to other techniques. Our technique, BlinkType, is based on blinking and allows users a fast input rate (at 13.47 wpm by participants using it for 45 minutes). As such, it is faster than other blink based techniques like BlinkWrite2 (5.3 wpm) and very competitive with other types of hands-free techniques.

\subsection{Eye Blinks and Properties}

Blinking is a bodily function of rapid, semiautomatic closing of the eyelids. In general, eye blinks can be classified as spontaneous blinks and reflex blinks \cite{bacher2004spontaneous}. Spontaneous blinking occurs without external stimuli and internal effort, and a person typically blinks 10-20 times per minute every 4-19 seconds \cite{leigh2015neurology}. Reflex blinks are activated by an external object touching to the cornea, bright light shining in front of the eye, or rapidly approaching objects. The duration of a single blink depends on the situations and could range from 100-400ms. Dennison et al. \cite{gurung2016mathematical} stated that blinking frequency is higher while wearing an HMD compared to a normal display monitor. This evidence is supported by a study from Kim et al. \cite{kim2005rescue} who suggested that the increased blinking frequency during VE immersion is due to side effects as well as visual fatigue and motion sickness. Following previous research on the use of eye blink outside of VR to support text entry, we have also explored it in pilot studies and found it promising. As such, we extended this research and developed a technique that accurately captures eye blinks and uses them for letter selection. As reported later, our technique leads to fast rate and is well-accepted by users.

\section{The Three Evaluated Hands-Free Text Entry Techniques}

We designed and implemented three hands-free text entry mechanisms, NeckType, DwellType and BlinkType, for text entry tasks in VR HMDs. We were interested in exploring whether letter selections with NeckType and BlinkType offered any advantages over DwellType. We used head rotation to control a cursor for all three techniques, while the same QWERTY virtual keyboard was used for all three techniques (see Figure 1). The keyboard was 370×156 pixels, which took about 14\% of the entire screen and was placed at the center of users’ view, both horizontally and vertically. It was within the 60 degree field-of-view in the HMD. We used the algorithm described in Goodman et al.’s Language modeling \cite{goodman2002language} for soft keyboards to predict key sequences. All three techniques used the same language model which combined the keypress model to predict the most probable words, especially in cases where the pointer hit the boundary of the keys. Additionally, we highlighted selected letters by enlarging them by 1.2 times and changing the background color as shown in Figure 1e.\par 
We also explored eye movement-related selection approaches such as gaze up or down. However, we excluded them due to some limitations: (1) eye movements could lead to involuntary typing while moving their gaze to browse letters; and (2) gazing down or up neededs to be confined within the target key area as otherwise it could produce selection errors. Thus, we excluded any eye movement-based selection techniques in our design.

\subsection{DwellType}

DwellType was a dwell-based text entry technique and allowed text selection by gazing and keeping the cursor on a target key for a period of time. Our DwellType technique was consistent with the other previous implementations of dwell techniques \cite{yu2017tap}. To enter a character, users first needed to move the cursor to the target key and kept it within the key area for 400ms. We used a circular progress bar to indicate dwell-time (Figure 1b), and a light beep sound to help users know whether the character was selected. We set the dwell time period based on prior research \cite{yu2017tap}, which has been also common among dwell-based techniques. If users wanted to enter the same character twice, the time needed is 800ms. Once the cursor moved to a new key, the dwell time would restart counting from zero. Users must move out from the keys to avoid typing the same letter continually. It is worth nothing that an adjustable dwell time could in theory possibly lead to higher text entry speed. However, dwell typing with adjustable dwell time has not been fully evaluated in VR, and as such, it is unknown how it will perform in virtual environments. The dwell time used was based on recent research \cite{yu2017tap} where 400ms was found to be ‘appropriate, under which users easily committed unintentional selections’. 

\subsection{NeckType}

NeckType leverages the forward (z-axis) acceleration movements, which can be accurately captured by the built-in IMU sensor of current HMDs, as the letter selection mechanism. NeckType is an enhanced version of DepthText \cite{lu2019depthtext, yu2019depthmove} and employed the Dynamic time warping (DTW) \cite{berndt1994using} algorithm for input recognition. To perform a selection with NeckType, users first moved the pointer on a target key with the head rotation and confirmed the character by moving the head forward for about 1cm, which the human neck can do with ease \cite{lu2019depthtext}.\par
We utilized DTW to recognize the z-axis input signals (see Figure 2). Although acceleration curves in the x-axis and y- axis did not show regular variation patterns, we recorded them simultaneously as well. To be able to use DTW, we collected neck movements that are comfortable and easy to do via a pilot study with 6 participants to produce the standard curve of DTW \cite{yan2018headgesture}. We conducted another pilot study collecting 5 users’ neck movements to determine the optimal thresholds for the system to detect a selection without any false positives. We found that the threshold in the z-axis was at most 0.78 while the thresholds in the x-axis and y-axis were at least 0.4 and 0.65 for the system to be able to recognize the selection accurately and effectively and, at the same time, to avoid unintentional selections while moving the cursor using head motions.

\begin{figure}[tb]
 \centering 
 \includegraphics[width=\columnwidth]{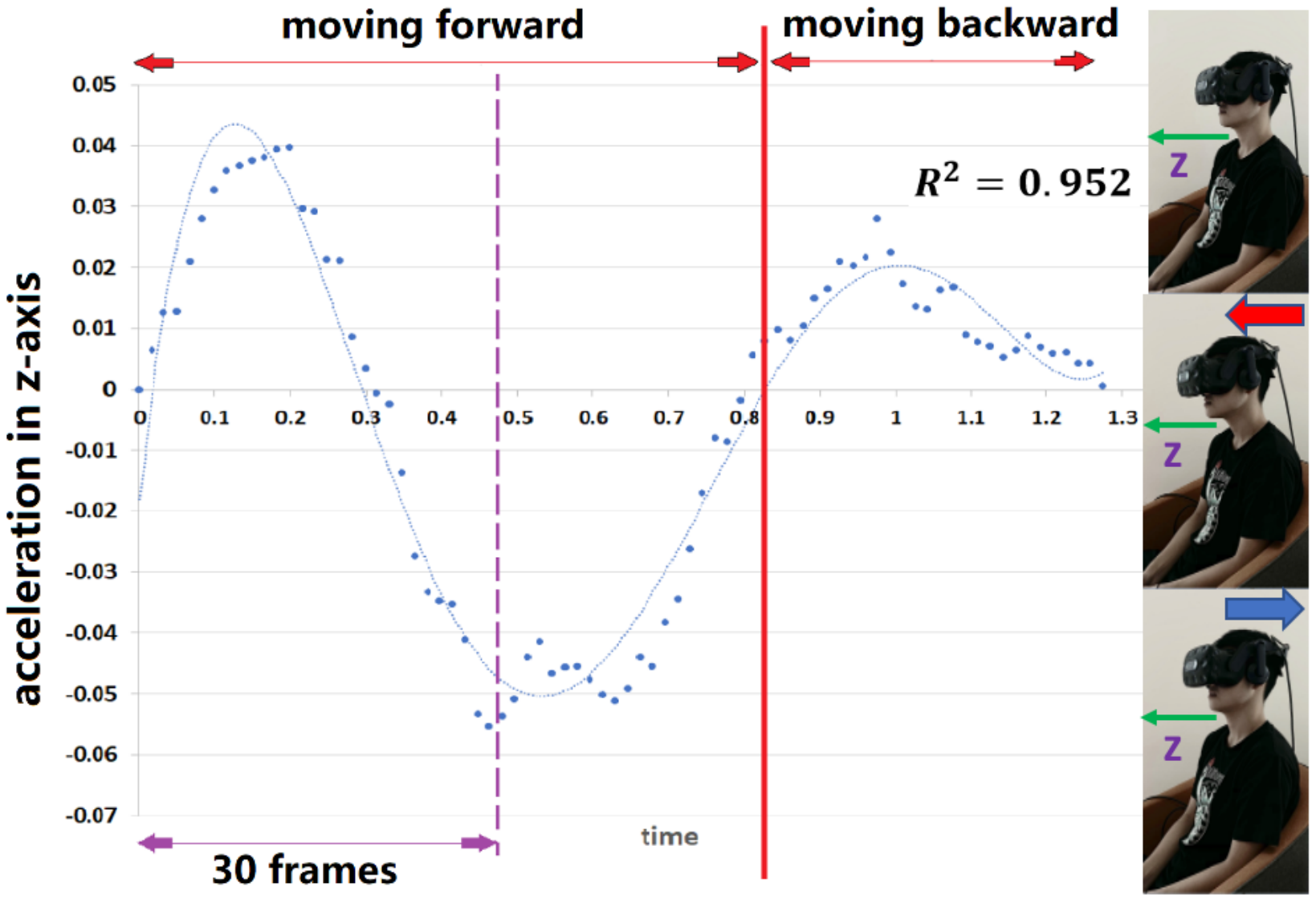}
 \caption{(left) Corresponding acceleration variation in the z-axis of a single head forward-backward motion. Such motion produces two sine-like waves (the first segment of the curve stands for moving forward and the second one for moving backward), which is not found in other axes. (right) A user is entering text using NeckType by moving the head forward in the VR system.}

\end{figure}

\subsection{BlinkType}

Eye blinks detection was achieved with Tobii eye trackers from the HTC VIVE Pro Eye. We also conducted a pilot study with 6 participants to explore suitable eye blink techniques (i.e., blinking with the left eye, blinking with the right eye, and blinking with both eyes) in a set of target selection tasks. Participants were required to select a total of 288 targets in both cardinal and intercardinal directions where results showed that the target selection accuracy was higher with both eyes blinking ($\sim$100\%) than left eye blinking (79.5\%) and right eye blinking (69.4\%). Many participants often accidentally close both eyes when they were supposed to perform tasks with single eye, leading to lower accuracy for the single eye conditions.\par
Consequently, blinking with both eyes was chosen for our BlinkType technique due to its nearly perfect accuracy. Moreover, to further reduce the possibility of false-positive selections, participatns were allowed to gaze at any space outside of the virtual keyboard to rest their eyes temporarily. To be consistent with the DwellType, 400ms was used as the minimum gap time between double-blink entries and so users could type double characters by closing their eyes for 800ms.\par
As started earlier, we used head rotations for controlling the cursor with all these three hands-free text entry techniques. There are other potential techniques such as gaze-based pointing that could be included in the study to control the cursor. However, we excluded gaze for pointing because in a pilot study we observed low accuracy with the gaze-based pointing compared to head rotations, especially when gaze pointing was used in conjunction with blinking and neck movements. In the end, we only used head rotation for moving the cursor over the letters. 

\section{User Study}
\subsection{Participants and Apparatus}

Thirty-six (36) participants who had no issue with neck and body movements and had normal or corrected-to-normal vision were recruited from a local university. A between-subjects design was used to evaluate the three hands-free techniques. There were 9 males and 3 females for NeckType and BlinkType with an average age of 19.7 and 19.8, respectively. For DwellType, there were 10 males and 2 females with an average age of 19.6. All participants were familiar with the QWERTY keyboard. None of them took part in the pilot studies that we conducted to find the suitable parameters for NeckType and BlinkType. Although the participants were not native English speakers, all of them were familiar with the English alphabet because English is the language of instruction at their university. The experiment was conducted on an HTC VIVE Pro Eye which had a 110 degree FOV. We used the controller only for users to proceed from one trial to another; otherwise, all three techniques were hands-free. The experimental application was developed with Unity3D.

\subsection{Experiment Design and Procedure}

We used a between-subjects design to avoid carry-over effects because the effect of motion sickness could accumulate overtime for VR systems. Participants’ demographic information was recorded with a questionnaire. At the beginning of the experiment, each participant was briefed with the details of text entry techniques and the VR HMD they would use. Participants then proceeded to transcribe phrases from the corpus in the practice stage ($\sim$10 minutes) to get familiar with the text entry technique and VR environment. After, they were asked to transcribe 48 phrases, which were divided into 6 sessions evenly. All phrases were randomly generated from the MacKenzie phrase set \cite{mackenzie2003phrase}. They were also instructed to enter the text as quickly and accurately as possible. Between sessions, we asked the participants to rest until they felt comfortable and ready to move to the next session. After the experiment, participants were asked to fill out the Simulator Sickness Questionnaire (SSQ) \cite{kennedy1993simulator} and the User Experience Questionnaire (UEQ) \cite{hinderks2018ueq}. We used SSQ and UEQ as they are commonly used to gather feedback from users in VR research \cite{wang2015evaluation, somrak2019estimating, kim2018virtual}. At the end, they were asked to share their experience and feedback about the text entry technique. The experiment lasted for about 50 minutes.

\subsection{Results}

To analyse text entry performance, we employed a two-way mixed ANOVA with Session (from session one to six) as the within-subjects variable and Technique (NeckType, DwellType and BlinkType) as the between-subjects variable. For the analysis of user experience feedback, we applied a one-way ANOVA with Technique as the between-subjects variable. Furthermore, Bonferroni correction was used for pairwise comparisons and Greenhouse-Geisser adjustment was used in case of violations of the sphericity assumption.\par
Text entry speed was measured in WPM, with a word defined as five consecutive letters, including spaces. The error rate is calculated based on standard typing metrics \cite{soukoreff2003metrics}, where the total error rate (TER) = not corrected error rate (NCER) + corrected error rate (CER). We report error rates based on TER and NCER.

\begin{figure}[tb]
 \centering 
 \includegraphics[width=\columnwidth]{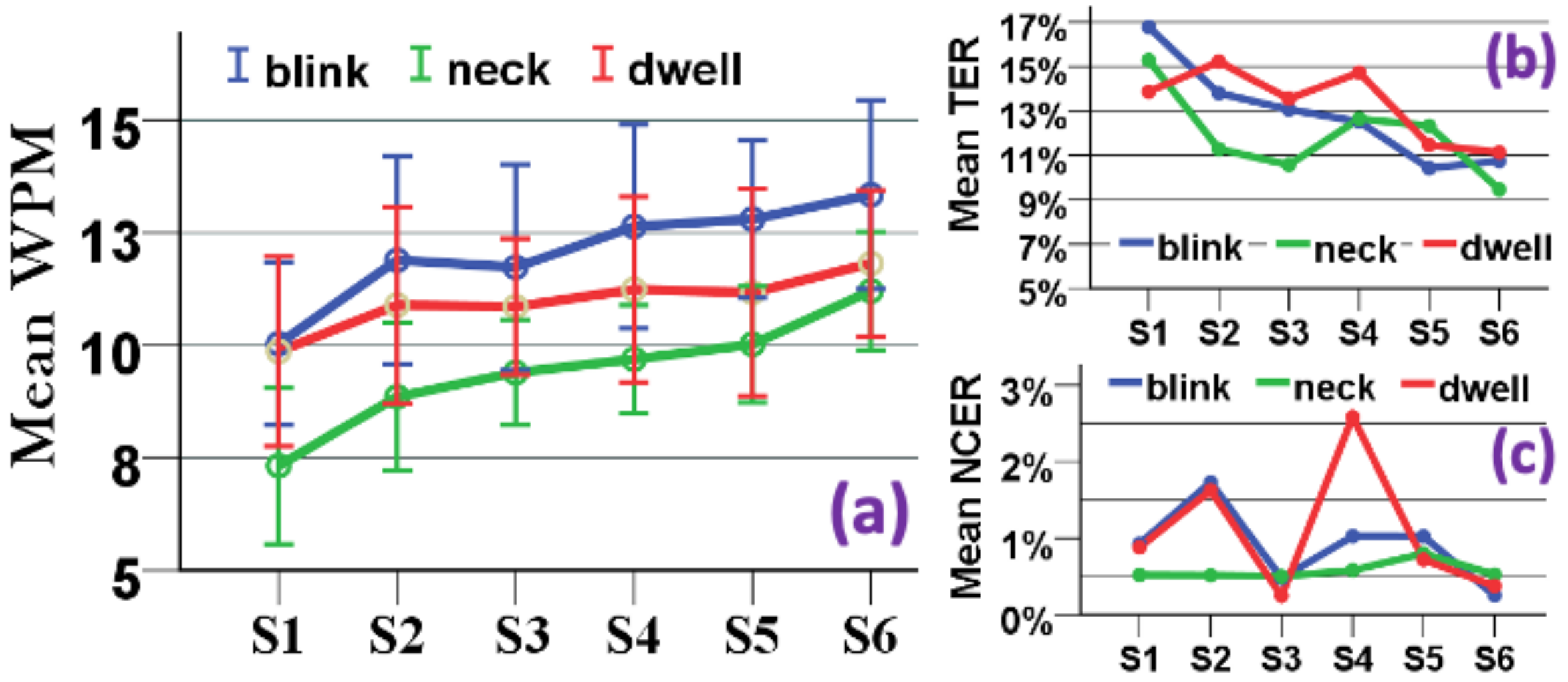}
 \caption{(a) The mean text entry speed of three hands-free input techniques through 6 sessions (‘S’ represents Session). Error bars represent ±1 standard error. (b) Mean total error rate (TER) and (c) not corrected error rate of three techniques through 6 sessions (‘S’ stands for Session).}

\end{figure}
\subsubsection{Text Entry Speed}
The two-way mixed ANOVA tests  yielded a significant effect of Session ($F_{5,165}=29.530, p<.001$) and Technique ($F_{2,33} = 3.716, p<.05$) on text entry speed. No significant interaction effect was found on Session × Technique ($F_{10,165} = 1.217, p>.05$). Post-hoc pairwise comparisons indicated significant differences between sessions 1-2, 1-3, 1-4, 1-5, 1- 6, 2-4, 2-5, 2-6, 3-6, 4-6 (all $p<.05$). In addition, post-hoc pairwise comparison found that BlinkType (M = 12.18, SD = 1.17) was significantly ($p<.05$) faster than NeckType (M = 9.43, SD = 0.41). Figure 3a shows the mean typing rate of the three techniques. BlinkType was faster, with an average speed of 13.47 wpm in the last session. NeckType and DwellType had similar typing rates (11.18 and 11.65 wpm, respectively). NeckType performed the lowest typing speed of 7.27 wpm in the first session but was improved to 11.18 wpm in the last session, with an increase of 53.78\%. Among the participants, the highest text entry speed appeared in session 4 of BlinkType (at the rate of 18.92 wpm).

\subsubsection{Total and Not Correct Error Rate}
The two-way mixed ANOVA tests revealed significant effects of Session on TER ($F_{5,165} = 3.960, p< .01$) and NCER ($F_{5,165} = 1.252, p<.05$). We did not find any significant effect of Technique on TER ($F_{2,33} = 0.181, p>.05$) and NCER ($F_{2,33} = 0.446, p>.05$). Figure 3b, 3c represents mean TER and NCER over 6 sessions of the three text entry techniques. With NeckType, TER dropped from the first session of 14.38\% to the sixth session of 9.04\%, with a decrease of 59.14\%. The mean TER in the last session with BlinkType and DwellType reached 10.44\% and 9.64\% respectively. The NCER showed irregular variation according to the data and reached at 1.01\%, 0.90\%, and 0.61\%, respectively, with NeckType, BlinkType and DwellType.

\begin{figure}[tb]
 \centering 
 \includegraphics[width=\columnwidth]{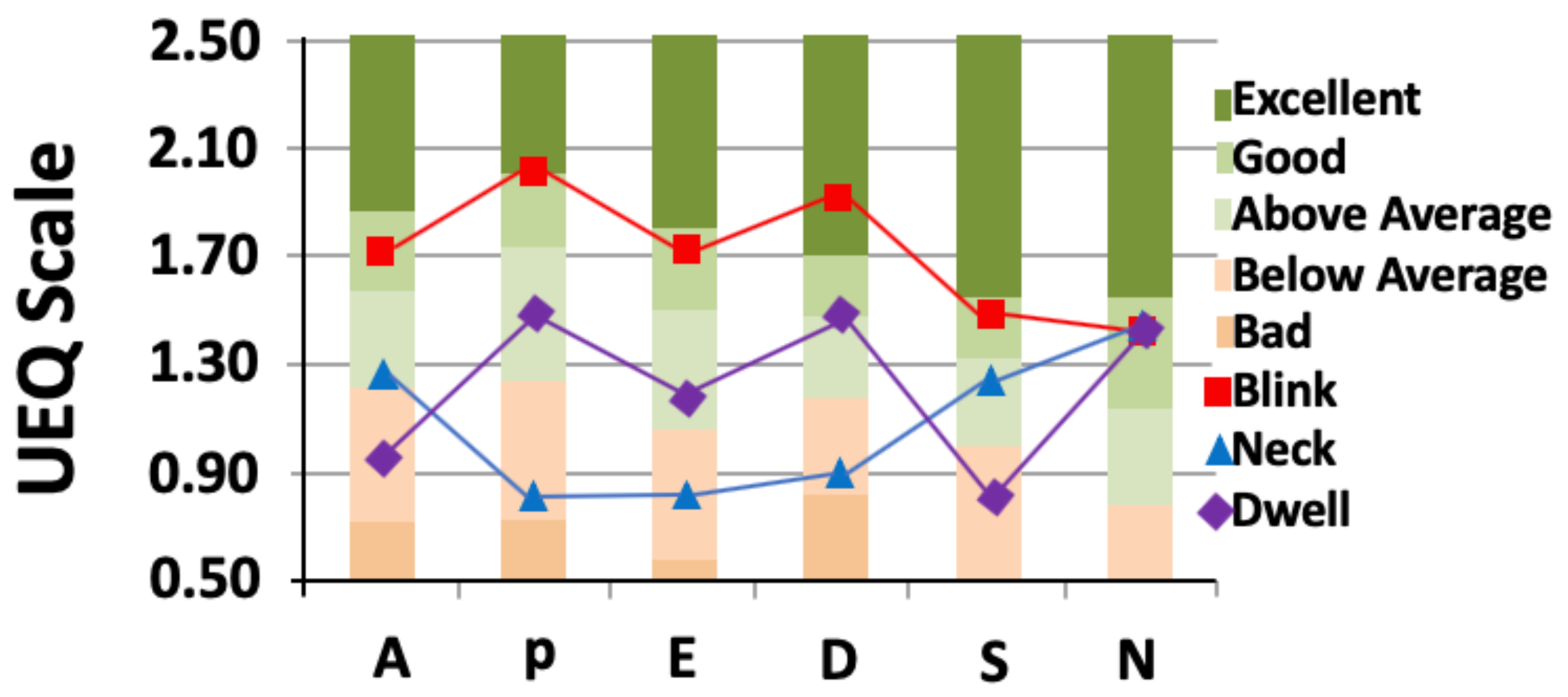}
 \caption{User Experience Questionnaire data regarding to its subscales: Attractiveness (A), Perspicuity (P), Efficiency (E), Dependability (D), Stimulation (S), and Novelty (N).}
\end{figure}

\begin{figure}[tb]
 \centering 
 \includegraphics[width=\columnwidth]{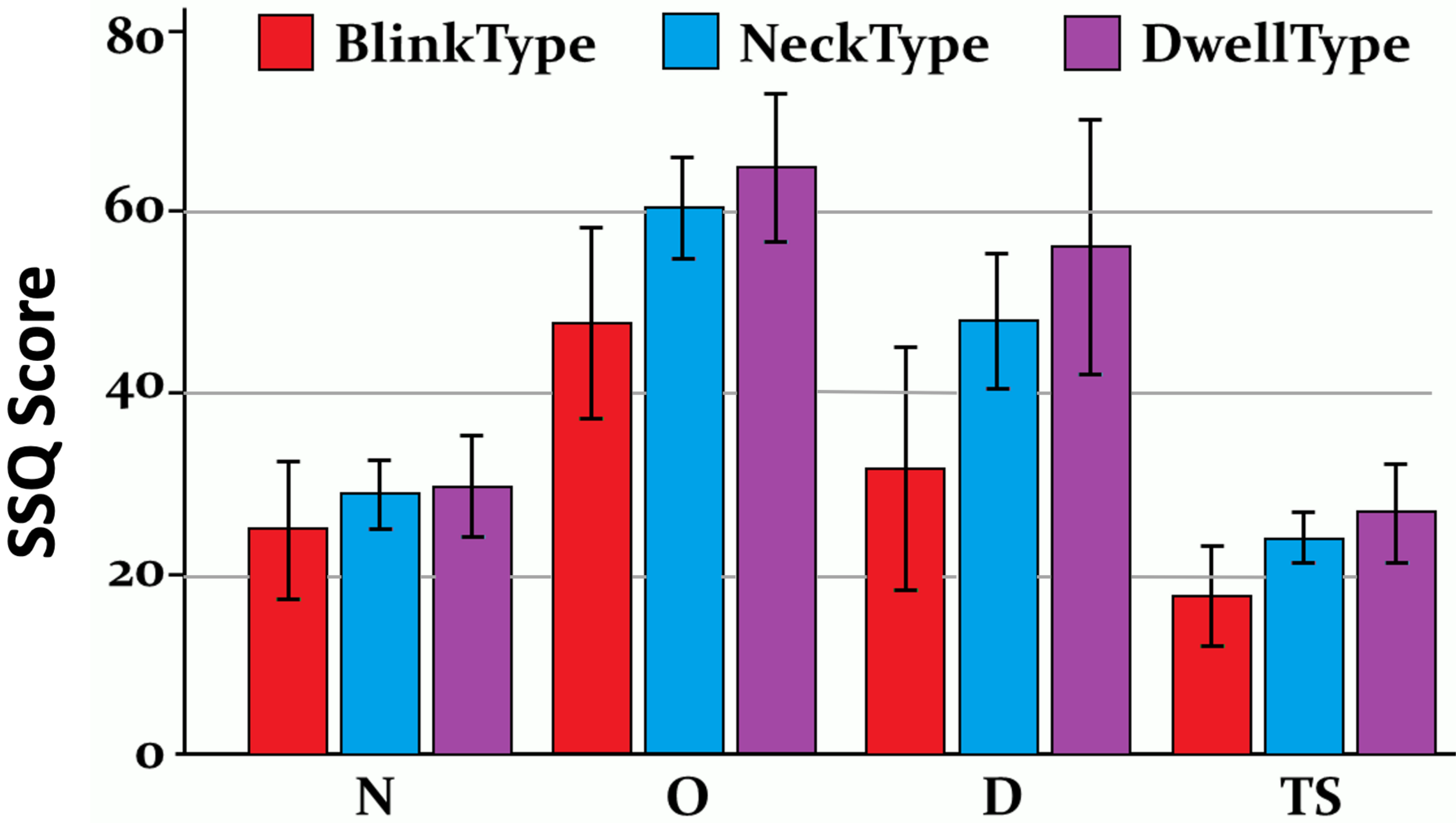}
 \caption{Simulator Sickness ratings in terms of Oculomotor (O), Nausea (N), Disorientation (D), and Total Severity (TS). Error bars represents ±1 standard error.}
\end{figure}

\subsubsection{User Experience Questionnaire (UEQ)}
The one-way ANOVA indicated that Technique had a significant ($F_{2,33} = 4.395, p<.05$) influence on Perspicuity. Post-hoc pairwise comparisons showed that the Perspicuity rating for BlinkType (M = 2.00, SD = 0.27) was significantly ($p<.05$) higher than NeckType (M = 0.81, SD = 0.17). The analysis also revealed that Technique had a significant ($F_{2,27} = 3.883,p<.05$) effect on Dependability. Post-hoc pairwise comparisons showed that the Dependability rating for BlinkType (M = 1.90, SD = 0.33) was significantly ($p<.05$) higher than NeckType (M = 0.88, SD = 0.13). No other significant effects were found. We performed a correlation analysis of Perspicuity and Dependability scores and wpm to understand why there were higher scores in performance. We found that Perspicuity (R = 0.354, $p<.05$) and Dependability (R = 0.436, $p<.05$) have statistically significant correlation on wpm. This implies that higher scores on Perspicuity and Dependability were associated with higher typing speed. Figure 4 depicts each UEQ subscale ratings for the three techniques. 

\subsubsection{Simulator Sickness Questionnaire (SSQ)}
Figure 5 shows the SSQ ratings for Oculomotor (O), Nausea (N), Disorientation (D), and Total Severity (TS). We could not observe any significant effect of Technique on N ($F_{2,35} = 0.187, p>.05$), O ($F_{2,35} = 1.134, p>.05$), D ($F_{2,35} = 1.072, p>.05$), TS ($F_{2,35} = 0.978, p>.05$).

\section{Discussion AND Future Work}
Our results show that BlinkType is a viable solution for hands-free text entry in VR. Users experience of BlinkType was rated Good to Excellent based on the UEQ data where DwellType and NeckType were only rated Below Average to Above Average (Figure 4). We believe that BlinkType is faster as it is not constrained by any dwell time to select or does not require neck movement in a forward-backward direction, which sometimes is a little cumbersome to perform. In addition, in short-term studies like this, it is often for participants’ text entry rate not to have peaked. Our results show this pattern also, where as can be seen in Figure 3a our participants’ rate has not peaked and may not likely to converge after six sessions. Additionally, the trend shows a gap between BlinkText and the other two, which could likely remain in place even if performance goes up further. Future work based on a longitudinal study will be useful to determine how long is required for users to reach peak entry rates and whether the gap between BlinkText and other hands-free techniques similar to DwellType and NeckType will remain in place.\par 
Though BlinkType is a promising solution for hands-free text entry, it has some limitations. For instance, BlinkType requires an eye tracker, which may not be available in all current VR HMDs. However, eye-tracking technology is becoming cheaper (e.g., IR LED can be a cheaper alternative to detect blinks on HMDs) and many VR manufacturers are now integrating eye-tracking solutions with their VR/AR HMDs—examples include HTC VIVE Pro Eye, Pico Neo 2 Eye, FOVE, HoloLens 2, Magic Leap 1, and LooxidVR. As such, eye tracking will likely be a standard feature of these device, and make BlinkType a very feasible technique.\par 
For all techniques, the average rate shows an increasing trend even in Session 6. This shows that after around 50 mins the fatigue factor was not significant. As stated earlier, our study did not look at long-term usage (days or weeks) of hands-free techniques on typing speed, which could be a future topic to explore. Related to this, we suspect that high values of oculomotor could have resulted from somewhat prolong use of HMDs and constantly searching for target keys. Being immersed in a virtual environment and doing gaze motions could cause oculomotor fatigue to some extent, especially for some users not experienced with HMDs; for example, Iskander et al. \cite{iskander2018review} indicate that motion is one cause of visual fatigue. In future, we would like to investigate fatigue issue in a long-term study.  Furthermore, we chose 400ms of dwell time based on a prior work \cite{majaranta2009fast} and was used by other researchers \cite{yu2017tap}. Exploring the effect of different dwell time (i.e., higher or lower) can be another possible avenue of the further work. In addition, data relevant to InterKeystroke Interval (IKI) analysis could have been recorded in future studies to provide further insights into the specific factors that contribute to performance of hands-free techniques in VR.\par
DwellType is currently a popular choice for hands-free text entry with VR HMDs. However, our participants expressed that they are often pushed to keep moving the cursor over the keyboard to go to the next character, which can be stressful and tiring. A similar concern has also been reported in \cite{sarcar2013eyek}. If users are not skilled at typing and are not familiar with the English alphabet, they might stop in a specific place for a while to avoid selecting unintended keys, which could further lead to fatigue and lower performance \cite{jacob1991use}. On the other hand, NeckType is an inexpensive hands-free text entry technique as it does not require any additional hardware for implementation. Additionally, it has the same text entry performance, UEQ, and SSQ as DwellType. Like BlinkType, with the increasing trend of speed (53.78\% improvement of text entry speed) as shown in Figure 3a, we expect participants could achieve a faster speed in NeckType than DwellType if further practice sessions are provided. Nevertheless, one shortcoming of NeckType is that users seem to need more training to be proficient (e.g., 10 minutes).\par
Based on our study results, we can extrapolate the following three recommendations for hands-free text entry in HMD VR:
\begin{itemize}
	\item [R.1] BlinkType is the preferred text entry technique in VR if eye-tracker is available. 
	\item [R.2] DwellType and NeckType can be used for typing if eye-tracker is not available.
	 \item [R.3] Designers should consider using both eyes’ blinking for triggering a selection as single-eye blink techniques lead to poor performance due to low accuracy.
\end{itemize}

\section{Conclusion}
In this paper, we report our exploration and evaluation of hands-free text entry techniques for virtual reality (VR) head-mounted displays (HMD). We compared three hands-free text entry techniques: BlinkType, NeckType and DwellType. Results for a user study with 36 participants showed that BlinkType offered the best performance, with participants being able to reach a mean typing speed of 13.47 wpm. With NeckType and DwellType, participants were able to type at the rate of 11.18 WPM and 11.65 WPM, respectively. Based on our study results, we suggest the use of BlinkType for hands-free text entry in VR, if eye tracking is available. Additionally, we found NeckType represents viable alternative approaches to dwell-based techniques that can offer users more control over the pace of selecting the characters and still lead to relatively fast entry.

\section*{Acknowledgments}
We want to thank our participants for their time and the reviewers
for their comments and suggestions that helped improve our paper. This research was supported in part by AI University Research Centre (AIURC) at Xi’an Jiaotong-Liverpool University (XJTLU), XJTLU Key
Program Special Fund (KSF-A-03 and KSF-02), and XJTLU Research
Development Fund.


\balance

\end{document}